\documentclass[english,preprint]{aastex}
\usepackage[T1]{fontenc}
\usepackage[latin1]{inputenc}
\usepackage{babel}
\usepackage{graphics}

\makeatletter

\providecommand{\LyX}{L\kern-.1667em\lower.25em\hbox{Y}\kern-.125emX\@}
\let\SF@@footnote\footnote
\def\footnote{\ifx\protect\@typeset@protect
    \expandafter\SF@@footnote
  \else
    \expandafter\SF@gobble@opt
  \fi
}
\expandafter\def\csname SF@gobble@opt \endcsname{\@ifnextchar[
  \SF@gobble@twobracket
  \@gobble
}
\edef\SF@gobble@opt{\noexpand\protect
  \expandafter\noexpand\csname SF@gobble@opt \endcsname}
\def\SF@gobble@twobracket[#1]#2{}

\makeatother
\begin{document}

\title{Is the Compact Source at the Center of Cas A Pulsed?}

\author{Stephen S. Murray, Scott M. Ransom, and Michael Juda}

\affil{Harvard-Smithsonian Center for Astrophysics, Cambridge, MA 02138}

\email{ssm@head-cfa.harvard.edu}

\author{Una Hwang}

\affil{NASA Goddard Space Flight Center, Greenbelt, MD 20771}

\author{Stephen S. Holt}

\affil{Olin College of Engineering, Needham, MA 02492}

\date{June 25, 2001}


\begin{abstract}
A 50 ksec observation of the Supernova Remnant Cas A was taken using
the \emph{Chandra} X-Ray Observatory High Resolution Camera (HRC)
to search for periodic signals from the compact source located near
the center. Using the HRC-S in imaging mode, problems with correctly
assigning times to events were overcome, allowing the period search
to be extended to higher frequencies than possible with previous observations
\cite{mur00a}. In an extensive analysis of the HRC data, several
possible candidate signals are found using various algorithms, including
advanced techniques developed by \cite{ran01} to search for low significance
periodic signals. Of the candidate periods, none is at a high enough
confidence level to be particularly favored over the rest. When combined
with other information, however (e.g., spectra, total energetics,
and the historical age of the remnant), a 12 ms candidate period seems
to be more physically plausible than the others, and we use it for
illustrative purposes in discussing the possible properties of a putative
neutron star in the remnant. We emphasize that this is not necessarily
the true period, and that a follow-up observation, scheduled for the
fall of 2001, is required.

A 50 ksec Advanced CCD Imaging Spectrometer (ACIS) observation was
taken, and analysis of these data for the central object shows that
the spectrum is consistent with several forms, and that the emitted
X-ray luminosity in the 0.1 -10 keV band is \( 10^{33}-10^{35}erg\, cm^{-2}sec^{-1} \)
depending on the spectral model and the interstellar absorption along
the line of sight to the source. The spectral results are consistent
with those of \cite{pav00} and \cite{cha01} .
\end{abstract}

\section{Introduction\label{sect:intro}}

The first high resolution X-ray image of Cas A was obtained using
the Einstein HRI. A search for a point source near the center of the
remnant resulted only in an upper limit on its luminosity (\cite{mur79, fab80}).
Other X-ray observations using ROSAT and ASCA also failed to detect
a central point source. This situation changed dramatically with the
first light image from the \emph{Chandra} X-ray Observatory taken
on 1999, August 20. In a relatively short (6ksec) Advanced CCD Imaging
Spectrometer (ACIS; \cite{gar97}) observation, the power of the \emph{Chandra}
high resolution X-ray telescope was demonstrated by revealing immediately
the existence of a point-like object near the center of the remnant
(\cite{tan99}). This object is likely to be either a neutron star
or black hole left over from the explosion of the progenitor star.
Subsequent searches in the radio (\cite{mcl01}) and optical (\cite{kap01, ryan01})
have yet to detect a point source at the center of Cas A. 

One distinguishing characteristic of a rotating neutron star would
be the presence of periodic pulses. The ACIS observations were made
in the timed exposure mode with a frame integration time of 3.24 seconds
so that searches for pulsations could be made only for fairly long
periods \( >10\, sec \). To search for shorter periods down to the
rotational breakup period for a neutron star (around 1 msec \cite{sha83}
), it was necessary to observe Cas A with the High Resolution Camera
(HRC; \cite{mur98, zom95} ). The HRC is an event triggered detector
with about \( 16\: \mu sec \) timing resolution. As part of the orbital
verification and calibration program, two 10 ksec observations of
Cas A were taken, one each with the HRC-I (imaging) and the HRC-S
(spectroscopy) detectors on 1999, September 3 and 5. When these and
additional public data were analyzed for pulsations from the point
source, no significant signals were found (\cite{cha01}). On average
there are only about 300 HRC counts per 10 ksec from the source (counts
within a 1 arc second radius), and Chakrabarty et al. placed an upper
limit of 35\% pulsed fraction for periods longer than 20msec.

Spectral analysis of the public ACIS data by both Chakrabarty et al.
and \cite{pav00} result in only marginal conclusions regarding the
nature of the central point source. {}``\emph{At present we do not
have a unique model to account for the observed properties of the
X-ray point source in Cas A}'' \cite{cha01}. {}``\emph{Critical
observations to elucidate the nature of CCO (Compact Central Object)
include searching for periodic and aperiodic variabilities, deep IR
imaging, and longer ACIS Chandra observations which would provide
more source quanta for the spectral analysis}'' \cite{pav00}. As
discussed below, we present a deeper observation of Cas A of 50 ksec
with the HRC for timing purposes, and 50 ksec with the ACIS for higher
accuracy spectral analysis than was possible from the calibration
data sets.

\subsection{Problems with HRC-I Timing}

Using the \emph{Chandra} X-Ray Observatory and the HRC, we obtained
a 50 ksec observation of Cas A specifically to search for pulsations
from the compact source detected near the center of the remnant. This
observation (OBSID 01505) was taken using the HRC-I on 1999 December
20. The standard \emph{Chandra} X-ray Center (CXC) pipeline processing
system was used to produce the Level 1 and Level 2 data products for
further analysis as described below. Data from the point source were
extracted from the event list in a \( \sim 1 \) arc second radius
region (8 HRC pixels) around the source. 

It was subsequently found that the HRC has a wiring error that incorrectly
assigns event times such that the assigned time is that for the previous
event trigger (\cite{mur00a, sew00}). If every event trigger resulted
in an event in the telemetry, this error could be easily corrected
by simply shifting event times by one event during ground processing.
However, due to telemetry limitations (\( 184\, events/sec \)) and
on-board event screening, not all event triggers necessarily result
in an event entering the telemetry stream. Therefore determining true
event times is not always possible, and under normal HRC operating
conditions cannot be done for a significant fraction of the events.

For OBSID 01505, the average rate of event triggers was about \( 300\, s^{-1} \)(total
event rate), many of which were due to cosmic rays and solar protons.
The HRC includes an active plastic scintillator coincidence shield
that is normally used to veto event triggers arising from penetrating
charged particles. With this veto enabled (normal mode), the average
rate of telemetered events was reduced to about \( 150\, s^{-1} \)(valid
event rate). Thus, in this normal operating mode, \( \sim 50\% \)
of the event triggers did not produce the events in the telemetry
which are needed to correctly assign proper event times to the events
coming from the central point source.

In order to evaluate the impact of the HRC timing error on our ability
to detect pulsations, we developed a high fidelity software simulation
of the detector and telemetry system. Simulations for this observation
(OBSID 01505) show that if no attempt is made to correct the timing
error, or if the only correction made is to shift the telemetered
time for each event by one event, then a sinusoidal pulse signal with
\( \preceq 20\% \) modulation amplitude or with a period of less
than 20 msec will be undetectable (similar to the conclusions of \cite{cha01}).

Some data recovery techniques have been investigated to improve the
probability that an event time will be correct. For example a filter
can be used to select event pairs where there is a high likelihood
that the events occurred in succession and therefore the time of the
first event of the pair can be obtained from the time assigned to
the second event. This class of filters reduces the number of events
that can be used for timing studies and therefore decreases sensitivity.
As discussed by \cite{ten01} in their analysis of a \emph{Chandra}
observation of the Crab Pulsar, this type of data recovery can be
useful when the pulse period is known, and therefore the signal needed
to make a measurement can be smaller than that needed for detection.
For OBSID 01505, no statistically significant period signal was found,
regardless of the algorithm or filter used.

\subsection{Solution using HRC-S in Imaging Mode}

For HRC-I using the coincidence shield and requiring anti-coincidence
between detector events and the shield for inclusion in the telemetry
stream reduces the cosmic ray background and limits the valid event
rate to a level below telemetry saturation. For HRC-S we would have
expected the same behavior; however, the coincidence window between
the event trigger and shield signal is too small, and an alternative
means of selecting events is used. The rate is reduced by pre-selecting
only about 1/2 of the detector active area using the edge blanking
feature of the instrument. This reduction is fine for spectroscopic
observations since the transmission gratings disperse the spectrum
along a thin region of the detector and the edges that are blanked
do not contain {}``interesting'' events. The consequences are essentially
the same as for the HRC-I. There are about 250 event triggers per
second, which would exceed the telemetry limit if all were allowed
to be processed as valid events. The on-board edge blanking veto selectively
suppresses {}``uninteresting'' events (for spectroscopy) to the
level where the telemetry is not filled and all processed events can
be transmitted to the ground. However, as with the HRC-I, the times
of events can not all be properly corrected since many event triggers
are not included in the data sent to the ground.

Fortunately, the HRC-S can be operated in a {}``special'' mode where
all event triggers result in events that are included in the telemetry.
In this mode, only the central MCP segment is able to initiate an
event trigger. This restriction reduces the background by about a
factor of three from the normal HRC-S rate (i.e., total event rate
goes from \( \sim 250\, c/s \) to \( \sim 90\, c/s \)). and therefore
allows the on-board event screening to be turned off. All event triggers
are processed as valid events and fit within the telemetry limit of
\( 184\, c/s \). This mode is designated as the HRC-S (Imaging) Mode.
It is now available for all observers and was used to re-observe those
AO-1 and AO-2 targets requiring the high time resolution of the HRC,
including our Cas A observation.

\section{Observations}

The new 50 ksec observation of Cas A (OBSID 01857) was carried out
on 2000 October 5 using the imaging mode of the HRC-S. The standard
Level 1 event list provided by the CXC pipeline processing system
was used as a starting point for more detailed analysis. This file
contains all of the HRC triggered events with their positions corrected
for instrumental (de-gap) and aspect (dither) effects. The events
were then time corrected by assigning the time of event n+1 to event
n, and then screened using the background reduction algorithm developed
by \cite{mur00b}.

A 50 ksec observation of Cas A on 2000 January 30 with the backside-illuminated
ACIS S-3 chip (OBSID 00114) provided roughly 5000 counts from the
point source. The CCD was operated in normal time exposure mode at
a temperature of \( -120^{\circ }C \). Instrument response files
for the spectral analysis were generated using software made available
by the CXC with calibration data from 2000 May.

\subsection{Image Analysis \label{sect:image}}

\begin{figure}[h!]
{\centering \includegraphics{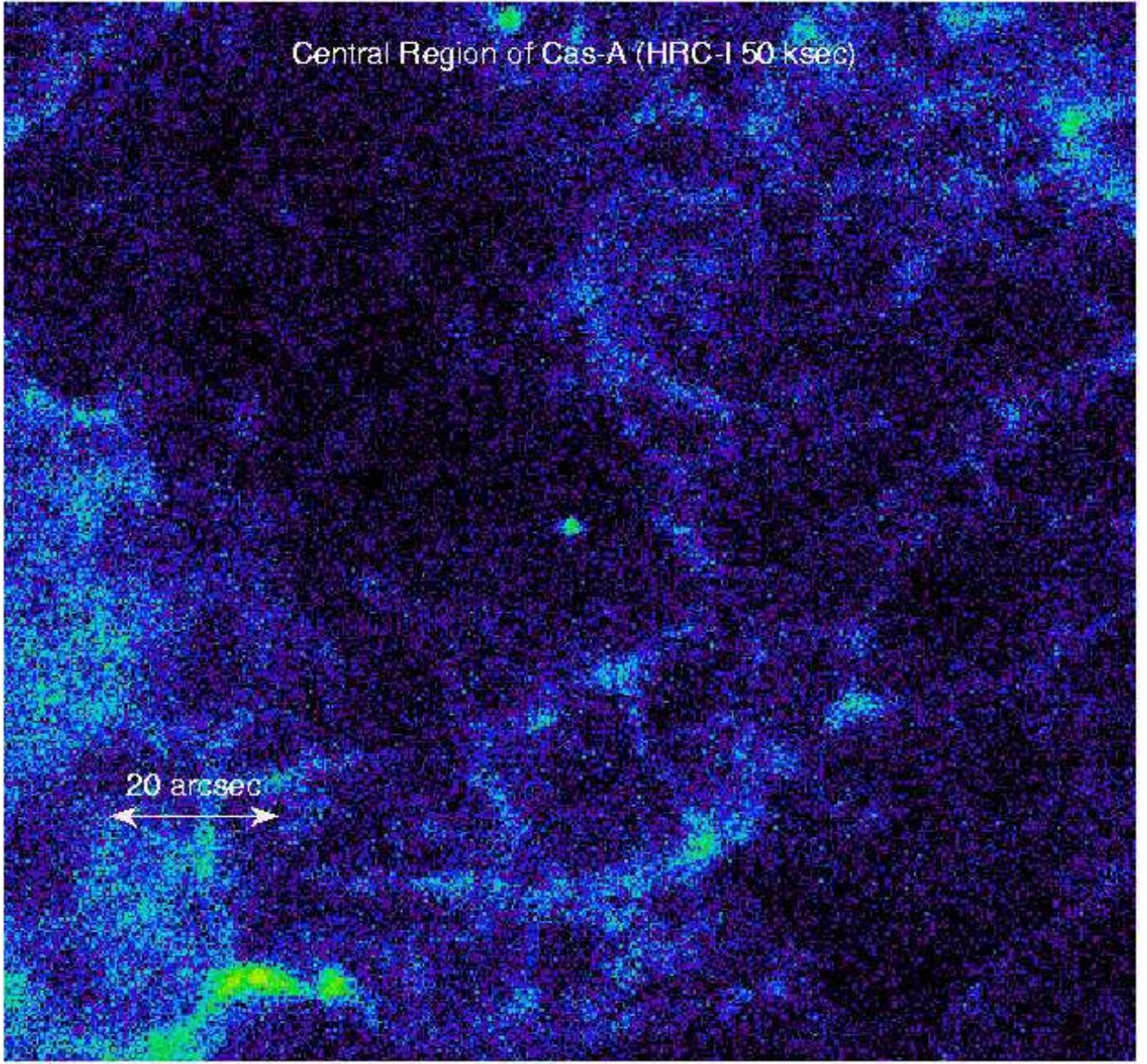} \par}

\caption{Central region of Cas A as observed in a 50 ksec image taken with
the HRC-I (OBSID 01505).\label{fig:hrcimage}}
\end{figure}

The images obtained with the HRC and ACIS provide upper limits to
the X-ray flux that might be associated with a synchrotron nebula
around the central compact object. Scaling from the Vela or Crab pulsar
nebula, we expect a nebula of order 10 arcsec in radius, and this
should emit a total luminosity of order 10 times that of the compact
object, assuming that it is a pulsar like the Crab. The compact source
in Cas A yields 1424 counts within 1 arc second radius (\( F_{ps}=2.85\times 10^{-2}HRC\: c/s \)
). Thus the nebula brightness should yield about 45 HRC counts per
square arc second in the 50 ksec observation or a total flux \( F_{neb}=2.85\times 10^{-1}HRC\: c/s \).\footnote{%
The area of the nebula is 100 times the region of emission from which
the pulsar flux is measured, but the total flux of the nebula is 10
times the pulsar's. Thus, the emission/arc second is about 10/100
as compared with the pulsar.
} The central region of Cas A is shown in Figure \ref{fig:hrcimage}.

For the extended region, excluding the 1 arc second radius region
around the compact object, we observe about 20 HRC counts per square
arc second, well below the expected value for a Crab-like nebula.
To further quantify this result, we examined the region around the
point source for excess emission using a wavelet decomposition (\cite{vik94})
to subtract the emission from the filamentary structures of the SNR
in the central part of the image as well as the point-like source
itself. We use the wavelet residuals as an estimate for the upper
limit from any nebula emission associated with the central object.
From the HRC-I image (OBSID 01505) we considered a central 2 arc minute
region for the wavelet decomposition. Using the residual image (i.e.,
the image after all of the wavelet components are subtracted), we
found a total flux of \( 1.4\times 10^{-1}HRC\: c/s \) within a 10
arc second radius (\( 3.3\times 10^{-1}HRC\: c/s \) within a 15 arc
second radius giving essentially the same surface brightness in both
cases, and indicating that most of the small scale remnant structures
have been removed), corresponding to an X-ray luminosity (within 10
arc seconds) of \( L_{neb}\leq 2.7\times 10^{34}ergs/sec\: (0.1-10.0\: keV) \)
assuming \( N_{H}=1.1\times 10^{22} \) and a photon spectral index
of 2.0. Using the ACIS-S data from OBSID 00114 we obtain a similar
result. We take this as an estimate of the upper limit to any emission
from a synchrotron nebula around the point source in Cas A.

\subsection{Timing Analysis\label{sect:time}}

In the image from OBSID 01857, the point source is clearly detected
at a location: \( 23^{h}23^{m}27^{s}.683, \) \( +58^{\circ }48^{'}43^{''}.21 \)
(CXO positions are accurate to within \( \pm 1 \) arcsec), within
1.7 arcsec of the position reported by \cite{tan99}. For the timing
analysis, only events within a \( \sim 1 \) arcsec radius of this
location were used (a total of 1424 events). The event times (provided
in terrestrial time by the CXC) were corrected to barycentric time
at the solar system barycenter using the definitive \emph{Chandra}
geocentric spacecraft ephemeris. The standard axBary program tool
(which uses the JPL DE450 solar system ephemeris) was provided by
the CXC yielding a time of arrival (TOA) list. 

We conducted initial searches for pulsations using two different methods,
a ``standard'' Fourier transform search, as well as a brute force
epoch folding based on the \cite{gl92} technique. In both cases,
provisions were made to eliminate losses in sensitivity due to finite
frequency resolution (i.e. ``scalloping'') and the possibility of
a signal with a slowly changing frequency (i.e. from pulsar spin-down).
The statistically most significant candidate in both searches was
a nearly sinusoidal signal with a period of 12.15 msec (82.267 Hz),
but numerous other plausible candidates were present as well. 

For the Fourier transform search, the events were binned into a 50
million point (1 msec time resolution) time series and then transformed
using an FFT (Fast Fourier Transform) algorithm. As the power spectrum
was dominated by white noise, all powers were normalized with the
number of events (1424), making the statistical properties of the
power spectrum equivalent to an exponential distribution with mean
and standard deviation of one. We utilized Fourier domain matched
filtering techniques to completely and coherently recover power that
had spread over up to \( \pm 4 \) frequency bins and to correct for
``scalloping'' effects for signals appearing at frequencies between
Fourier bins (\cite{ran01}). Such a search is warranted since young
pulsars are predicted to be rotating rapidly (\( \sim 10-100\: Hz \))
with a relatively large negative frequency derivative (\( -10^{-11}\: to\: -10^{-9}Hz/s \)).
An identical observation of the fundamental harmonic of the Crab pulsar
(\( f=29.9\: Hz \), \( \dot{f}=-3.77\times 10^{-10}Hz/s \)) would
show a drift of -0.94 frequency bins and power losses of from \( \sim 5-60\% \)
for uncorrected Fourier amplitudes (due to scalloping and frequency
drift). Incoherent summing of 1, 2, 4, and 8 harmonics was also used
to improve sensitivity to narrow pulse shapes. All frequencies and
periods that we mention will be the \emph{average} values over the
course of the observation.

The best candidate from the Fourier transform search had a frequency
\( f=82.267111(2)\: Hz \), or equivalently \( P=0.0121555260(3)\: s \),
and a marginally significant \( \dot{f}\sim -4(3)\times 10^{-10}Hz/s\: (\dot{P}\sim -7(5)\times 10^{-14}s/s) \).
The normalized power level was 19.5 which corresponds to \( <1.4\sigma  \)
when the number of independent trials searched is included. Subsequent
analyses searched a much wider range of frequency derivatives (\( \pm 100 \)
bin drifts) resulting in several candidates with higher significance
than the 12.15 msec candidate. The most significant was found as a
summation of 8 harmonics with the fundamental at \( f=41.472710,\: \dot{f}=-2.5(4)\times 10^{-9}Hz/s \),
and an overall significance of \( \sim 2.0\sigma  \) .

The modified folding search based on that found in \cite{gl92} has
several significant advantages over ``standard'' epoch-folding techniques.
The two most significant advantages are that the method does not depend
on a specific pulse shape and that it optimizes the signal-to-noise
for a profile based on the starting phase of the signal. These two
factors can increase the sensitivity of a search by a factor of a
few. We searched a frequency range of 0.001 to 500 Hz, and a frequency
derivative range of \( -5\times 10^{-10}\: to\: 5\times 10^{-10}Hz/s \).
Our frequency step size of \( 2.56\times 10^{-6}Hz \) is equivalent
to oversampling the Independent Fourier Spacing (\( IFS=1/T_{obs} \)
) by a factor of eight.

\begin{figure}[h!]
{\centering \includegraphics{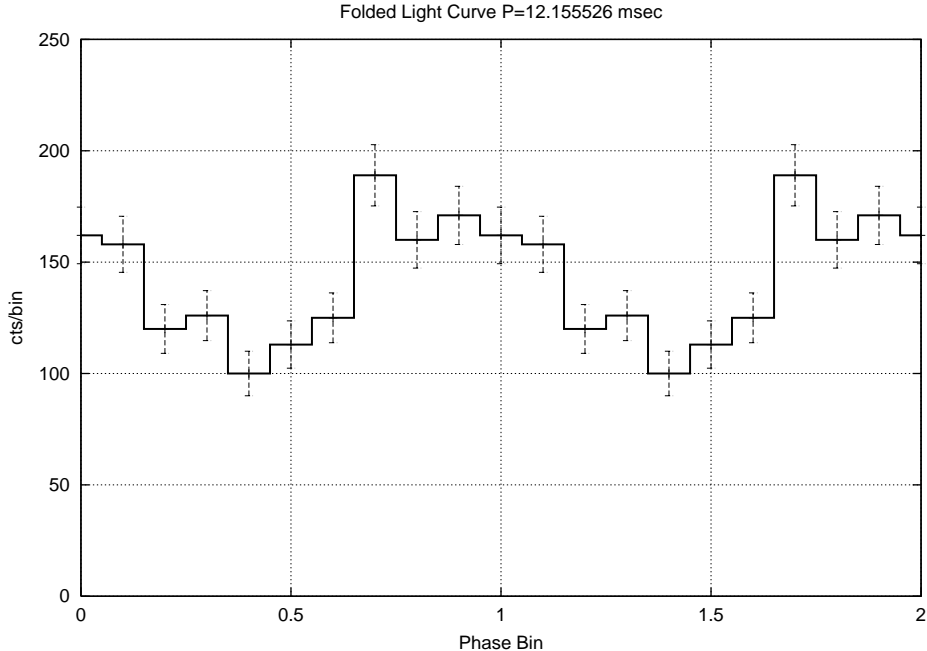} \par}

\caption{Epoch folding light curve for 12.155525 msec (82.267 Hz) candidate
period. Ten phase bins were used, and there are two cycles plotted,
the error bars are \protect\( 1\sigma \protect \) based on the counts
in each phase bin.  \label{fig:lc}}
\end{figure}

With our current implementation of the Gregory and Loredo search it
is very difficult to estimate an overall probability that a signal
is ``real'' when searching a very large frequency range. What we can
determine, however, is the relative merit of a candidate with respect
to another. Data containing a single strong signal would produce a
candidate list with the signal as the best candidate with the next
best candidate ranked significantly lower than the first. In our search,
the best candidate had \( f=82.267109(3)\: Hz \) with a marginally
significant \( \dot{f}\: \simeq -5(3)\times 10^{-10}Hz/s, \) consistent
with the most physically realistic candidate from the Fourier search,
and was ranked approximately 25 times better than the second best
candidate. The light curve for this candidate period is shown in Figure
\ref{fig:lc}, where we have used a simple epoch fold with a constant
period corresponding the frequency given above. The oscillation has
a modulation of \( \sim 25\% \) (peak - mean/ mean).

\subsubsection{Statistical Significance\label{sect:signif}}

We stress that although hard to quantify, the statistical significance
for the detection of period pulsations is low and cannot be taken
as definitive proof of their existence. However, the data do allow
us to place an upper limit on any sine-like oscillatory modulation
of the signal from the central source in Cas A.

For the HRC-S observation (OBSID 01857), we ran a standard epoch folding
analysis centered around the 12.155526 msec period found in the period
search discussed in Section \ref{sect:time}. The figure of merit
for the epoch fold is the reduced chi-squared (\( \chi ^{2}_{\nu } \))
obtained by comparing the folded (binned) data against the hypothesis
of a constant source. The period search was limited to 100,000 test
periods starting at 12.15 msec, and ending at 12.16 msec. The expected
period (12.155526 msec) was recovered with \( \chi ^{2}_{\nu }=6 \),
which (for 9 degrees of freedom) can be expected accidentally only
1 time in \( 10^{7} \). Given that \( 10^{5} \) trials were made
there is 1\% chance of obtaining this value for \( \chi ^{2}_{\nu } \)
from unpulsed data. As mentioned in Section \ref{sect:intro} we developed
a simulation program for the HRC and the \emph{Chandra} Observatory
that allows us to mimic the Cas A observations including the effects
of the wiring error in the instrument. We generated 100 simulated
data sets for the case of no pulsations, and ran each though the standard
epoch folding algorithm, testing \( 10^{5} \) periods per simulated
data set as described above. In total, \( 10^{7} \) epoch folds and
\( \chi ^{2}_{\nu } \) calculations were made. There was one occurrence
of \( \chi ^{2}_{\nu } \) exceeding 6, consistent with statistical
expectations. 

In an effort to determine a reasonable detection limit for a periodic
signal, we ran a series of simulations using a modulated signal, approximated
by a sinusoidal period of 12.155526 msec and with a modulation (defined
as the ratio of the peak deviation from the mean amplitude to the
mean amplitude) from 15\% to 35\%. That is, the number of counts at
phase bin \( i \) is given by \( c_{i}=A+B\sin (\phi _{i}) \), and
the modulation is \( m=\frac{B}{A} \). For each modulation value
we ran 100 simulations and obtained the average value for the figure
of merit (\( \left\langle \chi ^{2}_{\nu }\right\rangle  \)) and
its standard deviation. We conclude that if there is an oscillatory
signal from the central compact source, it is not highly modulated
(\( \leq 25-30\% \)) or we would have detected the signal with a
higher level of confidence than reported above. This result is consistent
with the conclusions of \cite{cha01} but is more restrictive in the
amplitude of the modulation and in the frequencies spanned.

\subsection{Spectral Analysis \label{sect:spectrum}}

Using the ACIS-S3 data from OBSID 00114, the source spectrum was extracted
from a 1.5 arcsec radius region centered on the point source, and
the background spectrum from an annulus surrounding the source region
with radii between 2.75 and 5.5 arcsec. We tried a number of different
background regions and found that the spectral analysis was not very
sensitive to the exact background region used. 

{
\begin{table}[h!]

\caption{X-Ray Spectral Fits to Cas A Central Point Source\label{tab:spectrum}}

\begin{tabular}{|c|c|c|c|c|c|c|}
\hline 
&
\( N_{H} \)&
Photon&
\( kT \)&
\( Radius_{bb} \)&
\( L_{d3.4} \)&
\( \chi ^{2} \)\\
Model&
(\( 10^{22}cm^{-2}) \) &
Index&
(keV)&
(km)&
(\( 10^{33}erg/s) \)&
\( \chi ^{2}/\upsilon  \)\\
\hline
\hline 
PL&
\( 2.23_{-0.12}^{+0.13} \)&
\( 4.13^{+0.17}_{-0.16} \)&
...&
...&
510&
161.5, 0.88\\
\hline 
ThBr&
\( 1.60^{+0.08}_{-0.09} \)&
...&
\( 1.15^{+0.07}_{-0.08} \)&
...&
5.9&
155.3, 0.84\\
\hline 
BB&
\( 1.05_{-0.07}^{+0.08} \)&
...&
\( 0.488^{+0.014}_{-0.015} \)&
0.46&
1.6&
206.2,1.12\\
\hline 
BB+PL&
\( 2.26^{+0.38}_{-0.27} \)&
\( 4.8^{+0.9}_{-0.6} \)&
\( 0.57^{+0.07}_{-0.07} \)&
0.23&
0.7, 2000&
148.1, 0.81\\
\hline 
BB+PL&
1.1 (fixed)&
\( 1.7^{+0.3}_{-0.4} \)&
\( 0.45^{+0.02}_{-0.01} \)&
0.54&
1.5, 0.4&
182.5,1.0\\
\hline
\end{tabular}\
\end{table}
\par}

\begin{figure}[h!]
{\centering \resizebox*{1\textwidth}{!}{\rotatebox{270}{\includegraphics{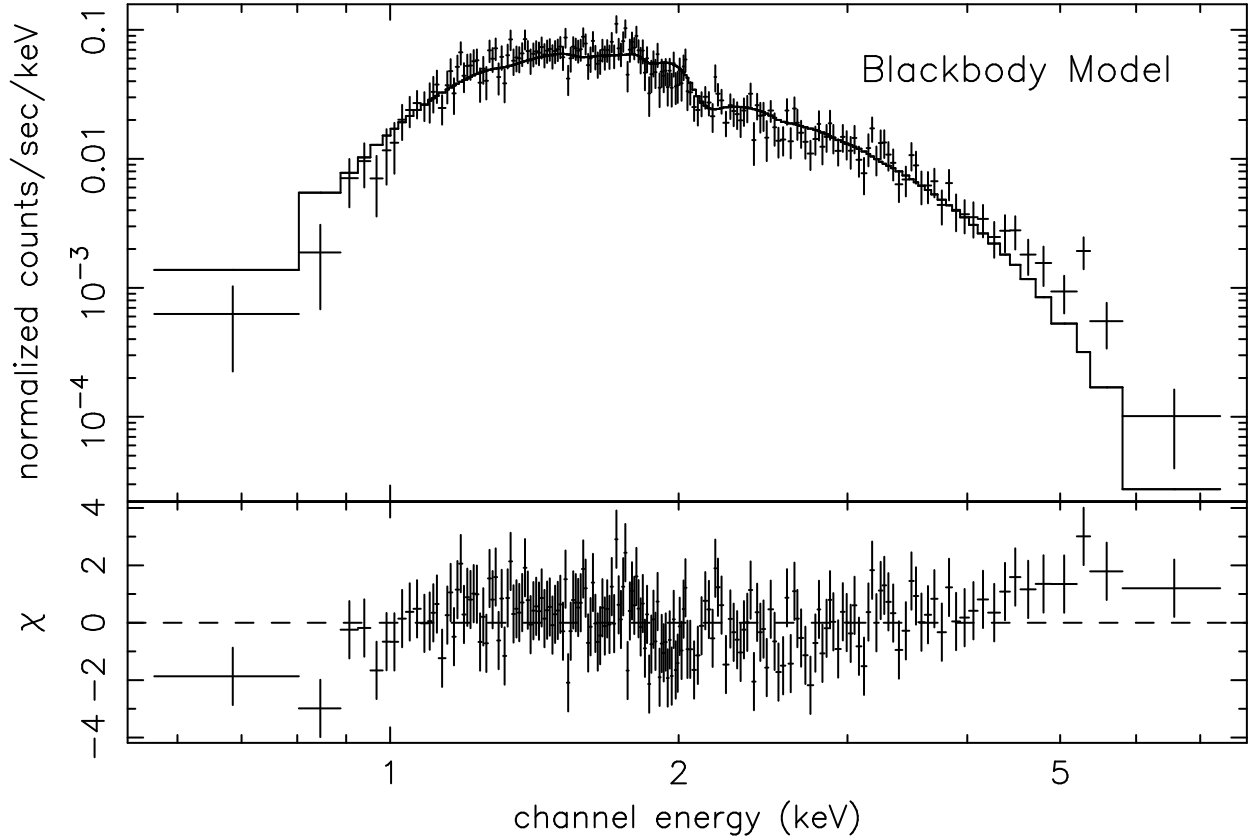}}} \par}

\caption{Spectral fit for the central source in Cas A. A simple blackbody
spectrum with interstellar absorption. \label{fig:spectrum}}
\end{figure}

The point source spectrum was fit to a number of simple spectral models,
including a blackbody, power-law, combination of blackbody and power-law,
and thermal bremsstrahlung (see \ref{tab:spectrum}). Figure \ref{fig:spectrum}
shows a representative spectral fit for a simple blackbody with interstellar
absorption. All the models gave acceptable fits with $\chi^2$ per
degree of freedom $\sim1.0$, but the thermal bremsstrahlung, power-law
and combination models required neutral Hydrogen column densities
that are in excess of the expected interstellar value of \( \sim 1\times 10^{22}\, cm^{-2} \)
in the direction of Cas A (\cite{keo98}). Unless there is significant
internal absorption, it would appear that these models are less favored.
For the power-law models, the slope is generally steep with a photon
index \( >4 \). In combination with a blackbody component with the
Hydrogen column density fixed at the expected value, the power-law
component is not unusually steep, but also does not dominate the flux.
The luminosity emitted at the source (corrected for interstellar absorption,
and assumed to be isotropic) varies between a few to several times
\( 10^{33}erg/s \), except for the models with a steep power-law
component, for which this luminosity is \( 5-20\times 10^{35}erg/s \).\footnote{%
Luminosities are 0.1-10.0 keV assuming a distance of 3.4 kpc for Cas-A,
except for the blackbody case where the bolometric luminosity at infinity
is given.
}

The spectral results are consistent with those based on the early
ACIS calibration observations reported by \cite{pav00} and \cite{cha01},
with the minor exception that we obtain a somewhat steeper index combined
with correspondingly higher absorption for the power-law fits where
\( N_{H} \) is unconstrained; this is most likely due to the improved
statistics of our longer observation. As found by those authors, the
power-law index is significantly steeper and the luminosity lower
than is normally found for X-ray pulsars. For the blackbody models,
the temperature is too high and the emitting area too small to correspond
to the entire surface of a neutron star. The emergent blackbody spectrum
will be modified by the atmosphere of the neutron star, however, in
a manner dependent on its composition, the strength of its magnetic
field, and orientation (e.g., see \cite{pav00}). In particular, the
true temperature can be significantly smaller than would be inferred
without accounting for the atmosphere. Even after including these
effects in their models, Pavlov et al. found that the emitting area
was still too large to be due to surface radiation from the entire
neutron star surface, and suggested that the emission is coming predominantly
from hot polar caps on the neutron star. The consistency of our spectral
parameters with those obtained by Pavlov et al. for the simple blackbody
models supports this conclusion.

\section{Interpretation\label{sect:interpret}}

As discussed above, the evidence for the existence of a pulsar in
Cas A is statistically weak, and the frequency of its possible period
and the degree of luminosity modulation are highly uncertain. \cite{pav00}
suggest that the source is a neutron star with hot spots. This model
is consistent with the spectral data and could result in an oscillatory
light curve like the one we observe, though our data imply only a
modest luminosity modulation, and such models will usually have a
strong modulation (\cite{cha01}). If our candidate 12 ms period is
correct, it would appear unfavorable for the interpretation of the
source as an anomalous X-ray pulsar, as suggested by \cite{cha01},
since these objects tend to have long periods on the order of seconds.
Likewise, it would tend to exclude neutron star accretion models with
high magnetic fields, since such models require slow rotations to
avoid centrifugal expulsion of the accreted material. Accretion onto
a low field neutron star model would still appear to be viable.

Without additional observational evidence, however, any interpretation
based on the existing data is speculative. A follow up \emph{Chandra}
observation of Cas A with the HRC-S (Timing) has been approved in
Cycle 2. It is scheduled for September 2001.

\subsection{Example Calculations\label{sect:calc}}

Despite the disclaimers cited above, we have taken what we consider
to be the most likely period (12.155526 msec) extracted from our observations
to calculate possible physical parameters for a putative Cas A pulsar.
If we assume that the Cas A pulsar is a classical magnetic dipole
(\cite{pac67,ost69}) and if we use the empirical relationship between
the nebula X-ray luminosity (\( L_{neb} \)) and the spin down energy
(\( \dot{E} \)) given by \cite{sew88} , or the relationship between
the pulsar luminosity and the spin down energy given by \cite{bec97},
then we can estimate the following properties for this object (we
use \( I\sim 1\times 10^{45}g\, cm^{2}s^{-2} \) and \( R=10\, km \)
for a nominal \( 1M_{\odot } \) neutron star):

\begin{enumerate}
\item We can place an upper limit to any synchrotron X-ray nebula based
on the HRC or ACIS images as described in Section \ref{sect:image}:
\( L_{neb}2.7\leq 10^{34}erg/s \)
\item Using the empirical relation \( \log L_{neb}=1.39\log \dot{E}-16.6 \),
we estimate \( \dot{E}\leq 5.2\times 10^{36}erg/s \) \cite{sew88}\\
Using the relationship \( L_{ns}=0.001\dot{E}, \) we estimate \( \dot{E}=3-5\times 10^{36}erg/s \)
(\cite{bec97})
\item From this estimate we calculate (taking \( \dot{E}=5.2\times 10^{36}erg/s \)
):\\
\[
\dot{P}=\frac{P^{3}\dot{E}}{4\pi ^{2}I}\leq 2.4\times 10^{-16}s/s\]
\\
\[
B=\sqrt{\frac{3c^{3}P^{4}\dot{E}}{32\pi ^{4}R^{6}}}\leq 5.4\times 10^{10}gauss\]

\item Using the historical age for Cas A \( \tau =320\, yrs \):\\
\[
\tau =\frac{P}{2\dot{P}}\left[ 1-\left( \frac{P_{0}}{P}\right) ^{2}\right] \Rightarrow P_{0}=0.01215\, s\]

\end{enumerate}
The above calculations are illustrative. They depend on the initial
assumptions regarding the validity of the classical magnetic dipole
model for the pulsar, and the estimated spin-down luminosity. The
lack of an observable X-ray synchrotron nebula suggests that the magnetic
field of the Cas A pulsar is considerably less then typical (e.g.,
as for the Crab), and is consistent with the above estimate. However,
without a confirmation of the pulse period, and a determination of
the pulsar spin down rate, additional interpretation and analysis
is too speculative to merit further discussion. We note that given
the precision with which the putative pulse period is measured (\( \sim 1\: nsec \)),
an observation of Cas A in \emph{Chandra} Cycle 2 will provide an
adequate baseline to measure \( \dot{P} \) to the limits suggested
by the above calculations.

\section{Conclusions\label{sect:conc}}

The point-like source near the center of Cas A has been observed for
50 ksec using the \emph{Chandra} X-ray Observatory HRC, operating
the HRC-S in a special mode to permit high resolution timing measurements.
We find weak evidence for a pulsed signal at a period of 12.155526
msec. The period-folded light curve is approximately sinusoidal and
the overall modulation is about 25\%. However, at approximately the
same significance level there are several period candidates, and we
cannot say at this time that the compact central object is indeed
a pulsar.

From a 50 ksec ACIS-S observation, the spectrum of the central compact
object in Cas A can be fitted to several simple models with more or
less equal significance, as in previously published results based
on shorter calibration observations. If the 12 ms period is real,
the central source probably has a low magnetic field, disfavoring
its interpretation as an anomalous X-ray pulsar. Despite the relatively
low modulation of the periodic signal, the X-ray emission may be due
to local hot spots near the polar caps of the putative neutron star
\cite{pav00}.

Assuming that the pulse period is correct, and taking a conservative
upper limit for an X-ray synchrotron nebula around the neutron star,
we find the surface magnetic field to be small (\( B\leq 5.4\times 10^{10}gauss \)
), and the spin down rate also to be small (\( \dot{P}\leq 2.4\times 10^{-16}s/s \)
). If confirmed in a follow up observation (\emph{Chandra} Cycle 2)
these values also imply that the pulsar was initially spinning with
a period close to 12 msec.

\acknowledgements{Acknowledgments}

The authors would like to thank Ramesh Nayaran and Irwin Shapiro for
helpful discussions and review of the manuscript. We also thank the
HRC Team for their efforts in solving the timing problems on the detector.
We appreciate the efforts of the \emph{Chandra} X-Ray Center in their
support for this observation. We acknowledge the use of the NASA Astrophysics
Data System in making it easier to review the literature and prepare
our reference list. This work was supported in part through NASA Contract
NAS 5-38248. Much of the timing analysis in this paper was carried
out on a Linux cluster at CfA funded by NSF grant PHY 9507695.

\end{document}